\documentclass[a4paper]{article}
\usepackage[utf8]{inputenc} 
\usepackage[T2A]{fontenc}
\usepackage{hyperref}
\usepackage{amssymb,amsmath}
\usepackage{floatrow}
\usepackage{layout}
\usepackage{amssymb} 
\usepackage{multirow}
\usepackage{caption}
\usepackage{authblk}
 \usepackage{indentfirst}
\usepackage{geometry}

\title{Conformal invariance, cosmological particle production and imitation of dark matter}
\author{V. A. Berezin$^1$ \and I. D. Ivanova$^2$}

\date{
	$^1$Institute for Nuclear Research of the Russian Academy of Sciences, \\
60th October Anniversary Prospect 7a, 117312 Moscow, Russia \\ \texttt{berezin@inr.ac.ru}\\%
	$^2$Institute for Nuclear Research of the Russian Academy of Sciences, \\
60th October Anniversary Prospect 7a, 117312 Moscow, Russia \\ \texttt{pc\textunderscore mouse@mail.ru}\\[2ex]%
}

\begin{document}
\maketitle

\begin{abstract}
Using a model for an ideal fluid with a variable number of particles, a phenomenological description of the processes of particle production in strong external fields is investigated. The conformal invariance of the creation law is shown, which imposes rather rigorous restrictions on the possible types of sources. It appears that the combinations with the particle number density can imitate dark matter within this model.
\end{abstract}

\section{Introduction}

\par Additional symmetry increases the probability of the emergence of the Universe from ``nothing'' \cite{Vilenkin}.  We assume that the conformal invariance is the fundamental symmetry we sought after. This idea is supported, among others, by Roger Penrose \cite{Penrose} and Gerard 'tHooft \cite{Hooft}.
\par We use the Lagrangian for the perfect fluid in Eulerian variables proposed by J. R. Ray \cite{Ray} modified in such a way that the rate of the particle production enters explicitly \cite{Berezin}. This method allows us to describe particle production in the presence of strong external fields phenomenologically at the classical level but with the back reaction taken into account. It is shown that conformal ivariance of the total action leads to the case where we are dealing actually with some sort of the Sakharov’s induced gravity \cite{Sakharov}. 
\par It appears that the particle creation law is itself conformally invariant. Assuming that particles are produced by the scalar field we get rather rigorous restrictions on the possible types of sources. They include conformally invariant combinations of the geometric quantities and scalar field as well as the particle number density. To our great surprise, it appears that it is the combinations with the particle number density that give the contribution to the hydrodynamical part of the total energy-momentum tensor and act like the dust. They can be interpreted as the dark matter. We would like to emphasize that it is not the real matter but the echo of the quantum process of the particle creation.

\section{Local conformal transformation}

\par Riemannian geometry is completely determined by the metric tensor $g_{\mu\nu}$ . The affine connection
$\Gamma^\lambda_{\mu\nu}(x)$ is Levi-Civita connection in this case, 
\begin{equation} \label{gamma}
\Gamma _{\mu \nu }^{\sigma }=\Gamma _{\nu \mu }^{\sigma }, \quad g_{\mu \nu ; \sigma} =0,\quad \Gamma _{\mu \nu }^{\sigma }=\frac{1}{2}g^{\sigma \lambda }\left ( g_{\mu \lambda ,\nu }+g_{\nu \lambda ,\mu }-g_{\mu \nu ,\lambda } \right ),
\end{equation}
it defines the parallel transport of vectors and tensors and their covariant derivatives
\begin{equation} \label{nabla}
l^\mu_{; \lambda}= l^\mu_{\;,\lambda} +\Gamma_{\lambda\nu}^\mu \, l^\nu ,
\end{equation}
where ``comma'' denotes a partial derivative while ``semicolon'' denotes covariant derivative .

The Riemann tensor $R^{\mu}_{\phantom{\mu}\nu\lambda\sigma}$ is defined as follows:
\begin{equation} \label{curv}
	R^{\mu}_{\phantom{\mu}\nu\lambda\sigma}=\frac{\partial \Gamma^\mu_{\nu\sigma}}{\partial x^\lambda}-\frac{\partial \Gamma^\mu_{\nu\lambda}}{\partial x^\sigma}+\Gamma^\mu_{\varkappa\lambda}\Gamma^\varkappa_{\nu\sigma}-\Gamma^\mu_{\varkappa\sigma}\Gamma^\varkappa_{\nu\lambda},
\end{equation}
Ricci tensor $R_{\mu\nu}$ is its convolution,
\begin{equation} \label{Ricci}
	R_{\mu\nu}=R^{\lambda}_{\phantom{\mu}\mu\lambda\nu}.
\end{equation}
The curvature scalar is $R=g^{\mu\nu}R_{\mu\nu}$.
\par Local conformal transformation
\begin{equation} \label{local}
ds^2=\Omega^2(x)d\hat s^2=\Omega^2(x)\hat g_{\mu\nu}dx^\mu dx^\nu,
\end{equation}
does not change the coordinates. Here $\Omega(x)$ is the conformal factor, and we denote by ``hats'' the conformally  transformed quantities. The metric and its determinant are transformed, evidently, in the following way:
$$
g_{\mu \nu }=\Omega ^{2} \, \widehat{g}_{\mu \nu },\quad g^{\mu \nu }=\frac{1}{\Omega ^{2}}\, \widehat{g}^{\, \mu \nu }, \quad \sqrt{-g}=\Omega ^{4}\, \sqrt{-\widehat{g}} .
$$
\par Later we will also need the Weyl tensor $C_{\mu \nu \lambda \sigma }$, which is the traceless part of $R_{\mu \nu \lambda \sigma }$,

\begin{multline} \nonumber
C_{\mu \nu \lambda \sigma }=R_{\mu \nu \lambda \sigma }-\frac{1}{2}R_{\mu \lambda }\, g_{\nu \sigma }+\frac{1}{2}R_{\mu \sigma  }\, g_{\nu \lambda  }-\\-\frac{1}{2}R_{\nu \sigma }\, g_{\mu \lambda  }+\frac{1}{2}R_{\nu \lambda }\, g_{\mu \sigma }+\frac{1}{6}R\left ( g_{\mu \lambda }\, g_{\nu \sigma }-g_{\mu \sigma }\, g_{\lambda \nu } \right ) .
\end{multline}
Its most peculiar property is the conformal invariance: $C^{\mu }_{ \; \nu \lambda \sigma }=\hat{C}^{\mu }_{\; \nu \lambda \sigma }$.

\section{Phenomenological description of particle creation}

In what follows we will consider the perfect fluid as the matter field  and choose for its action integral the following one \cite{Ray},
\begin{eqnarray} \label{R3d}
S_{\rm m}&=& -\!\int\!\varepsilon(X,n)\sqrt{-g}\,d^4x + \int\!\lambda_0(u_\mu u^\mu-1)\sqrt{-g}\,d^4x
\nonumber \\ 
&&+\int\!\lambda_1(n u^\mu)_{;\mu}\sqrt{-g}\,d^4x + \int\!\lambda_2 X_{,\mu}u^\mu\sqrt{-g}\,d^4x.
\end{eqnarray}
The dynamical variables are the particle number density $n(x)$, the four-velocity $u^\mu(x)$ and the auxiliary variable $X(x)$.

The energy density $\varepsilon$ provides us with the equation of state $p=p(\varepsilon)$, where
\begin{equation} \label{2Ga}
p=n\frac{\partial\varepsilon}{\partial n} -\varepsilon 
\end{equation}
is the hydrodynamical pressure.

The variations of $S_{\rm m}$ in Lagrange
multipliers $\lambda_0$, $\lambda_1$ and $\lambda_2$ impose constraints, the four velocity normalization $u^\mu u_\mu=1$, the particle number conservation $(nu^\mu)_{;\mu}=0$ and the enumeration of trajectories $X_{,\mu}u^\mu=0$, respectively. 

The energy momentum tensor equals
\begin{equation} \label{T}
T^{\mu\nu}=(\varepsilon+p)u^\mu u^\nu -p g^{\mu\nu}.
\end{equation}

\par The article \cite{Berezin} shows that the process of particle creation can be described at the phenomenological level by modifying the corresponding constraint in the perfect fluid Lagrangian:
\begin{equation} \label{Phiinv}
(nu^\mu)_{;\mu}=\Phi(\rm inv),
\end{equation}
where ``the creation law'' $\Phi$ depends on the invariants of the fields responsible for the creation process. 

Let us show that the left-hand side of the creation law multiplied by $\sqrt{-g}$ is conformally invariant,
\begin{equation} \label{nhat}
n=\frac{\hat n}{\Omega^3},  \quad u^\mu=\frac{\hat u^\mu}{\Omega},  \quad \sqrt{-g}=\Omega^4 \sqrt{-\hat g},
\end{equation}
hence
\begin{eqnarray} \label{numu2}
(n u^\mu)_{;\mu}&=&\frac{1}{\sqrt{-g}}(n u^\mu\sqrt{-g})_{,\mu}= \frac{1}{\sqrt{-g}}\left(\frac{\hat n}{\Omega^3} \frac{\hat u^\mu}{\Omega}\Omega^4\sqrt{-\hat g}\right)_{,\mu}=
\nonumber \\ 
&=& \frac{1}{\sqrt{-g}}(\hat n\hat  u^\mu\sqrt{-\hat g})_{,\mu},
\end{eqnarray}
and we obtain, as a consequence, that $\Phi \sqrt{-g}$ is conformal invariant. 

\par Without external classical fields the particle are created solely by the vacuum fluctuations due to the gravitational field, therefore $\Phi$ should depend on the geometric invariants. If we restrict ourselves to invariants that are quadratic in the curvature tensor, the only conformally invariant combination for Riemannian geometry in the four-dimensional case is the square of the Weyl tensor $C^{2}=C_{\mu \nu \lambda \sigma }\, C^{\mu \nu \lambda \sigma }$. If some external scalar field $\varphi$ contributes to the process of particle production the following addition to the creation law may be suggested: 
$$
\varphi \square \varphi -\frac{1}{6}\, \varphi ^{2}\, R+\Lambda \, \varphi ^{4},
$$
since it is invariant under a conformal transformation when the scalar field changes as:
\begin{equation}
    \varphi=\frac{\hat{\varphi}}{\Omega} \,.
\end{equation}
Here $\square$ is Laplace–Beltrami operator.
\par Our particles are the on-shell quanta of the scalar field. Therefore, they also can produce "new" particles. The rate of such production should depend on the number density of the "old" particles, i.e. it is some function of n. The most natural choices are: $\varphi \, n$ and $n^{\frac{4}{3}}$, both of them, being multiplied by $\sqrt{-g}$, form conformal invariants. Thus, our creation law becomes:
$$
\Phi=\alpha \, C^2+\beta\, \left( \varphi \square \varphi -\frac{1}{6}\, \varphi ^{2}\, R+\Lambda \, \varphi ^{4} \right)+\gamma _{1}\, \varphi \, n+\gamma _{2}\, n^{\frac{4}{3}} .
$$
 
\section{Induced gravity}
\par The matter action integral $S_m$ reads now as follows:
\begin{eqnarray} \nonumber
S_{\rm m} =  -\!\int\!\varepsilon(X,\varphi,n)\sqrt{-g}\,d^4x + \int\!\lambda_0(u_\mu u^\mu-1)\sqrt{-g}\,d^4x+
\nonumber \\ 
+\int\!\lambda_1 \left ( (n u^\mu)_{;\mu}-\Phi \right )\sqrt{-g}\,d^4x + \int\!\lambda_2 X_{,\mu}u^\mu\sqrt{-g}\,d^4x,
\end{eqnarray}
Note that $\varepsilon=\varepsilon(X,\varphi,n)$.
\par \quad If one demands that the gravity itself is conformal invariant, then one needs nothing more. This is because the Lagrange multiplier $\lambda_1$ is defined, actually, up to a constant. We adopt this point of view and get an example of Sakharov's induced gravity \cite{Sakharov}. Thus, we assume:
$$
S_m=S_{tot} \, .
$$
\par Evidently,
$$
\frac{\delta S_{tot}}{\delta \Omega }=\frac{\delta S_{m}}{\delta \Omega }=0,
$$
on the solutions. Nevertheless, it is very instructive to make use of the specific structure of our matter action integral. It is not difficult to show that the only part of $S_m$ that matters, is
$$
\!\int\!\varepsilon(X,\varphi,n)\sqrt{-g}\,d^4x .
$$
\par Remembering that $n=\frac{\hat{n}}{\Omega ^{3}}, \quad \varphi =\frac{\widehat{\varphi }}{\Omega }, \quad \sqrt{-g}=\Omega ^{4}\, \sqrt{-\widehat{g}}$, one gets:
\begin{equation} \label{eq}
\varphi\,  \frac{\partial \varepsilon }{\partial \varphi }+3n\, \frac{\partial \varepsilon }{\partial n}=4\, \varepsilon,    
\end{equation}
with the solution:
\begin{equation} \label{F}
\varepsilon =F\left ( \frac{n}{\varphi ^{3}} \right )\, \varphi ^{4},    
\end{equation}
where F is an arbitrary function of one variable.
\par There are two important examples. The first is dust matter with $p=0$:
$$
\varepsilon= \mu_0 \, n\,  \varphi.
$$
\par The second is radiation for which $\varepsilon=3p$ and therefore $\varepsilon=\nu_0 \, n^{\frac{4}{3}}$ does not depend on $\varphi$. Note the resemblance with two "hydrodynamical" terms in the creation law.

\section{Equations of motion and  constraints}

\par Let us derive the (modified) hydrodynamical equations of motion and corresponding energy-momentum tensor.
\begin{eqnarray} \nonumber
S_{\rm m} =  -\!\int\!\varepsilon(X,\varphi,n)\sqrt{-g}\,d^4x + \int\!\lambda_0(u_\mu u^\mu-1)\sqrt{-g}\,d^4x+
\nonumber \\ 
+\int\!\lambda_1 \left ( (n u^\mu)_{;\mu}-\gamma _{1}\, \varphi \, n-\gamma _{2}\, n^{\frac{4}{3}}+...\right )\sqrt{-g}\,d^4x+
\nonumber \\ + \int\!\lambda_2 X_{,\mu}u^\mu\sqrt{-g}\,d^4x .
\end{eqnarray}
\par Dynamical variables are n, $u^{\mu}$, $\varphi$ and X:
\begin{equation} \label{phi}
 \delta \varphi : \quad   \beta \, \left ( \lambda _{1} \square \varphi +\square \left ( \lambda _{1}  \varphi \right )+4\lambda  _{1}\, \Lambda \varphi ^{3 }-\frac{1}{3}\lambda _{1}\, \varphi\,  R \right )+\gamma _{1}\,  n=-\frac{\partial \varepsilon }{\partial \varphi },   
\end{equation}
\begin{equation} \label{n}
\delta n: \quad -\frac{\partial \varepsilon}{\partial n }-\lambda _{1, \sigma}\, u^{\sigma}-\lambda _{1}\gamma _{1}\, \varphi -\frac{4}{3}\lambda _{1}\gamma _{2}\, n^{\frac{1}{3}} = 0,    
\end{equation}
\begin{equation} \label{u}
\delta u^{\mu}: \quad \lambda_{2}\, X_{,\mu }+2 \lambda _{0}\, u_{\mu }- \lambda_{1,\mu }\, n =0,   
\end{equation}
\begin{equation} \label{X}
\delta X: \quad -\frac{\partial \varepsilon }{\partial X}-\left ( \lambda  _{2}\, u^{\sigma } \right )_{;\sigma }=0     
\end{equation}

\par The corresponding constraints are:
\begin{equation} \label{0}
\delta \lambda_{0}: \quad u_{\sigma}\, u^{\sigma}-1=0,    
\end{equation}
\begin{equation} \label{1}
\delta \lambda_{1}: \quad \left ( nu^{\sigma } \right )_{;\sigma }= \Phi,    
\end{equation}
\begin{equation} \label{2}
\delta \lambda_{2}: \quad X_{,\sigma} \, u^{\sigma}=0.    
\end{equation}
From the equation (\ref{u}) multiplied by $u^{\mu}$ and constraints we get:
\begin{equation} \label{lambda0}
 2\lambda _{0}=-n\, \frac{\partial \varepsilon}{\partial n }-\lambda _{1}\gamma _{1}\, \varphi\, n -\frac{4}{3}\lambda _{1}\gamma _{2}\, n^{\frac{4}{3}}.   
\end{equation}

\par Let's calculate the hydrodynamical part of the energy-momentum tensor that is the energy-momentum tensor of the perfect fluid plus contribution from the $\gamma_1$ and $\gamma_2$ terms. From the general definition:
$$
S_{m}=-\frac{1}{2}\int T^{\mu \nu }\, \delta g_{\mu \nu }\, \sqrt{-g}\, d^{4}x,
$$
taking into account the equation (\ref{lambda0}), we get:
\begin{multline} \label{hydro}
T^{\mu \nu }_{hydro}=\varepsilon \, g^{\mu \nu }-2\lambda _{0}\, u^{\mu }\, u^{\nu }+g^{\mu \nu }\, \left ( n\, \lambda _{1, \sigma}\, u^{\sigma}+\lambda _{1}\gamma _{1}\, \varphi\, n+\lambda _{1}\gamma _{2}\, n^{\frac{4}{3}} \right )=\\=\left ( \varepsilon +p+\lambda _{1}\gamma _{1}\, \varphi\, n+\frac{4}{3}\, \lambda _{1}\gamma _{2}\, n^{\frac{4}{3}} \right )\, u^{\mu }\, u^{\nu }- g^{\mu \nu }\, \left ( p+\frac{1}{3}\, \lambda _{1}\gamma _{2}\, n^{\frac{4}{3}} \right ).    
\end{multline}
The remaining parts of the momentum energy tensor are:
\begin{multline}
T^{ \mu \nu }[\varphi ]=\lambda _{1} \beta \Lambda\, \varphi ^{4}\, g^{\mu \nu } -\beta \, \partial _{\sigma }\left ( \lambda _{1}\varphi \right )\partial ^{\sigma }\varphi\, g^{\mu \nu }+\beta \, \partial ^{\mu }\left ( \lambda _{1}\varphi  \right )\partial ^{\nu }\varphi+\\+\beta \, \partial ^{\nu }\left ( \lambda _{1}\varphi \right )\partial ^{\mu }\varphi  +\frac{\beta }{3}\left \{ \lambda _{1}\varphi ^{2} \, G^{\mu \nu }-\triangledown^{\mu } \triangledown^{\nu } \left ( \lambda _{1} \varphi ^{2 }\right )+g^{\mu \nu }\, \square \left ( \lambda _{1} \varphi ^{2} \right )\right \},\\
T^{ \mu \nu }[C^{2} ]=-8\alpha  \, \left ( \triangledown _{\sigma }\, \triangledown_{\eta }+\frac{1}{2} \, R_{\sigma \eta } \right )\left ( \lambda_{1}\, C^{\mu \sigma \nu \eta } \right ),
\end{multline}
where $G^{\mu \nu }$ is Einstein tensor.
Since we are dealing with induced gravity, then:
$$
T^{\mu \nu }=T^{\mu \nu }_{hydro}+T^{ \mu \nu }[\varphi ]+T^{ \mu \nu }[C^{2} ]=0.
$$
\par There appeared two accompanying persons: $\gamma_1$ - dust-like $(\lambda_1>0)$ and $\gamma_2$ - radiation-like $(\lambda_1>0)$. They are not the real ones because the particle number density n refers to the real created particles whose equation of state can be arbitrary ( whatever you like ). Thus, they are something like the echoes of the creation process itself. The best name for them is "gravitating mirages".  

\par Finding a general solution to the equations of motion is quite a complex task, so we will limit ourselves to considering two special cases: $\lambda _{1}=const$ and $\varphi=0$. In the first case from the equation (\ref{n}) we get: 
$$
\frac{\partial \varepsilon }{\partial n}=-\lambda _{1}\, \left ( \gamma _{1}\, \varphi +\gamma _{2}\, n^{\frac{1}{3}} \right ),
$$
the solution is:
$$
\varepsilon =-\lambda _{1}\, \left ( \gamma _{1}\, n\varphi +\gamma _{2}\, n^{\frac{4}{3}} \right )+f(\varphi ).
$$
Function $f(\varphi)$ then can be found from the relation (\ref{F}):  
$$
f(\varphi )=C \varphi ^{4},
$$
where C is arbitrary constant. The hydrodynamical part of the energy-momentum tensor is: $T^{\mu \nu }_{hydro}=C\,\varphi ^{4}\, g^{\mu \nu }$, it means that in this case the term $f(\varphi)$ in $\varepsilon$ is equivalent to the shift of the constant $\Lambda$. The equation of motion for $\varphi$ reduces to the following:
\begin{equation} \label{phi1}
2\lambda _{1}\beta \left ( \square \varphi -\frac{1}{6}R\, \varphi+2\Lambda \, \varphi ^{3} \right )+4C\varphi ^{3}=0.   \end{equation}
\par The conformal invariance of the equations of motion and the creation law makes it possible to simplify the problem by fixing the gauge. In the gauge $\varphi=\varphi_0=const$ from the equation (\ref{phi1}) we get:
$$
R=\frac{12\varphi_{0} ^{2}}{\beta \lambda _{1}}\, \left (C+\lambda _{1}\, \beta \Lambda  \right )=const,
$$
therefore the space-time in question is equivalent to the geometry with constant scalar curvature up to a conformal factor. 
\par In the second case equation (\ref{eq}) implies that our perfect fluid is radiation then according to (\ref{phi}) either $n$ or $\gamma_1$ is zero. If $\gamma_1=0$ it follows from the (\ref{n}) and (\ref{hydro}) that: 
$$
\lambda _{1, \sigma}\, u^{\sigma}=-\frac{4}{3}\, n^{\frac{1}{3}}\left ( \nu _{0}+ \lambda _{1} \, \gamma _{2}\right ),
$$
$$
T^{\mu \nu }_{hydro}=\frac{1}{3}\, n^{\frac{4}{3}}\left ( \nu _{0}+ \lambda _{1} \, \gamma _{2}\right )\, \left ( 4\, u^{\mu }u^{\nu }-g^{\mu \nu } \right ).
$$
Using the gauge $n=n_0=const$ and the comоving coordinate system where $u^{\sigma}=\delta _{0}^{\sigma }$ we can find $\lambda_1$ considering that it depends only on the proper time $t$:
$$
\lambda _{1}(t)=-\frac{\nu _{0}}{\gamma _{2}}+\left (\lambda _{1}(0)+\frac{\nu _{0}}{\gamma _{2}}  \right )\, exp\left \{ -\frac{4}{3}\, \gamma _{2}\, n_{0}^{\frac{1}{3}}\,  t\right \}.
$$
Note that $\lambda _{1}$ tends to a constant $-\frac{\nu _{0}}{\gamma _{2}}$ while $t \to \infty $ if $\gamma_2>0$.

\section{Discussion and conclusions}

\par The main focus of our conclusions is an interpretation of $\gamma_1$ and $\gamma_2$ terms. In cosmology $\gamma_1$ - term becomes the dark matter. It is dark because it is not the matter at all, but the "memory" of the process of the real particle creation. The conditions for its existence is $\lambda_1>0$ and $n \neq 0\, (>0)$. Therefore, the real particles should already be produced. The dark matter will exist even after the particle creation stops.   
\par In the same way, the $\gamma_2$ - term becomes the hot universe, even without real photons and the real temperature.    
\par Both of them are just images, but they are gravitating! Thus, we get two \textit{smiles of Cheshire cat}.

\end{document}